\documentclass[a4paper]{spicawStyle}
\usepackage{graphicx,natbib}

\newcommand{\cii}{{\rm C}\,{\sc ii}}
\newcommand{\oi}{{\rm O}\,{\sc i}}
\newcommand{\LIR}{L_{\rm IR}}

\begin{document}

\title{Cooling lines as probes of the formation and
buildup of galaxies and black holes}

\author{P.P.~van der Werf\inst{1} \and M.~Spaans\inst{2}} 

\institute{
Leiden Observatory, Leiden University, P.O.~Box 9513, NL - 2300 RA Leiden,
The Netherlands
\and
Kapteyn Astronomical Institute, Groningen University, P.O.~Box 800,
NL - 9700 AV, Groningen, The Netherlands}

\maketitle 

\begin{abstract}
We discuss the use of SPICA to study the cosmic history of star formation and
accretion by supermassive black holes. The cooling lines, in particular the
high-$J$ rotational lines
of CO, provide a clear-cut and unique diagnostic for separating the
contributions of star formation and AGN accretion to the total infrared
luminosity of active, gas-rich galaxies. We briefly review existing efforts
for studying high-$J$ CO emission from galaxies at low and high redshift.
We finally comment on
the detectability of cooling radiation from primordial (very low metallicity)
galaxies containing an accreting supermassive black hole with SPICA/SAFARI.
\keywords{Galaxies: formation -- Galaxies: nuclei -- Missions: SPICA}
\end{abstract}

\section{Introduction}

For several decades, our only direct probes of the high-redshift
universe were QSOs and radio galaxies. While significant insight
into the properties and evolution of these populations was obtained, almost
nothing was known about the population of normal galaxies at high $z$. This
has changed radically over the last decade. Deep imaging and spectroscopy 
from the ground and from space have revealed a rich and diverse population of
high-$z$ galaxies, and enormous progress has been made in establishing a
complete inventory of the high-$z$ universe, in terms of mass and metallicity
budget, energy output and evolution over cosmic time. In addition, first
investigations of the genesis of the fundamental scaling relations of galaxies
(e.g., the fundamental plane and
the Tully-Fisher relation) are being undertaken,
and will provide insight into the origin of the galaxy population itself.

One of the most remarkable scaling relations is the $M_\bullet-\sigma$
relation, i.e., the relation between the mass of a nuclear
supermassive black hole (SMBH) and the velocity dispersion of the
spheroid (elliptical or bulge) hosting the black hole
\citep{Magorrianetal98,FerrareseMerritt00}. The fact that the black
hole mass and total galaxy mass are closely related is truly
remarkable, given that there is a factor of $\sim10^8$ between the AU-size
Schwarzschild radius of the black hole and the kpc-size dimension of
the galaxy. This is generally interpreted as evidence that the
formation and growth of the SMBH is directly related to the formation
process of the stellar population, e.g., in a violent burst of star
formation. The picture of the simultaneous build-up of a SMBH with an
extreme burst of star formation provides a new context for the
luminous and ultraluminous infrared galaxies (LIRGs and ULIRGs), long
suggested to be the birthplaces of SMBHs powering active galactic
nuclei (AGNs). Studies of local ULIRGs demonstrate that they plausibly
evolve into moderate mass field ellipticals \citep{Genzeletal01}.
The giant ellipticals in the local universe would then be the
result of similar but much more extreme events at higher
redshifts. Indeed, the population of submillimetre
galaxies (SMGs) provides the more luminous high redshift counterpart of the
local LIRGs and ULIRGs and may be responsible for the formation of
local massive spheroids and for generating QSO activity at high
redshift \citep{Blainetal99}.

\begin{figure}[th]
  \begin{center}
    \includegraphics[width=8cm]{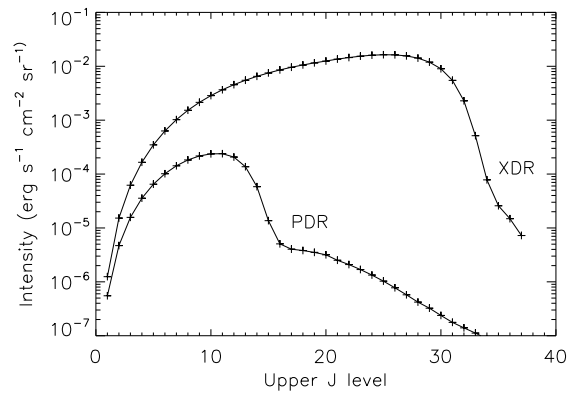}
  \end{center}
  \caption{CO emergent intensity distributions for a PDR and an XDR
with {\it identical\/} impinging energy density, as described in
Section~\ref{vanderWerfP.SecDiag}.
Note that
high-$J$ CO emission is expected in both cases, but vastly more prominently
in the XDR \citep{SpaansMeijerink08}.}
\label{vanderWerfP_fig1}
\end{figure}

\section{Star formation history and accretion history}
\label{vanderWerfP.SecDiag}

Theoretically, a number of mechanisms have been proposed for
establishing the $M_\bullet-\sigma$ relation \citep{Murrayetal05,Croton06},
but attempts at observational tests of these
models are extremely challenging and have sometimes produced
conflicting results \citep{Shieldsetal06,Alexanderetal05}.

SPICA provides a new window on this problem with its sensitivity in the
far-infrared region. This region contains the key cooling transitions of
photon-dominated regions (PDRs, heated by stellar UV radiation) and X-ray
dominated regions (XDRs, heated by the X-ray radiation from an accreting SMBH).
Advanced PDR/XDR models \citep{SpaansMeijerink08} reveal the significantly
different spectral patterns from XDRs and PDRs. In both cases cooling is
provided by the principal fine-structure lines ([\cii] $158\,\mu$m, [\oi]
63 and
$145\,\mu$m with in some regions of parameter space contributions from
[Si\,{\sc ii}] and [Fe\,{\sc ii}] lines) as well as rotational lines of
CO and rotational and
vibrational lines of H$_2$. However, the flux distribution over the CO
rotational ladder is vastly different for PDRs and XDRs.
In the UV, a 10\,eV photon penetrates a column of about
$10^{21}\,$cm$^{-2}$ (for solar metallicity) before being absorbed by
dust grains. UV photons also have the right energy to dissociate CO
between 11.2 en 13.6\,eV\null.
In contrast, a 1\,keV X-ray photon traverses a column of
$10^{22}\,$cm$^{-2}$ before absorption by hydrogen and metals occurs, since
the absorption cross section scales as $\sim$1/energy$^3$. Because
of their high energy, neither X-rays nor the induced secondary electrons
destroy CO efficiently.
Furthermore, UV photons cause photoelectric heating when electrons
are liberated from (small) dust grains, with an efficiency of
no more than 1\%. X-rays on the other hand
heat through ionization and (mostly) the subsequent
Coulomb interactions of liberated very fast electrons, with
an efficiency of about $10-50$\% (depending on the electron abundance).
Consequently, PDRs, associated with star
formation, contain CO layers that are relatively cool ($10-100\,$K), while
XDRs, associated with black hole accretion, 
are characterized by deeper layers of much warmer CO ($20-500$\,K).

In Fig.~\ref{vanderWerfP_fig1} a comparison is presented between a PDR and an XDR
for a density of $n=10^5$\,cm$^{-3}$ and an illuminating radiation
field of $160\,$erg\,s$^{-1}$\,cm$^{-2}$, representative of
a region like the Orion Bar in the Milky Way (in the UV) or gas at a
hundred parsecs from an accreting black hole powering a Seyfert nucleus
with a luminosity of $10^{44}$ erg s$^{-1}$ (in X-rays).
It should be noted that the PDR case included here pushes the most 
extreme range expected in starburst galaxies,
while X-ray fluxes as low as 5\,erg\,s$^{-1}$\,cm$^{-2}$,
impinging on $\sim10^5$\,cm$^{-3}$ gas are still capable of producing
substantial CO emission at very high $J$ CO levels. The CO ladder at 
high $J$ levels thus constitutes 
a unique diagnostic of SMBH accretion activity in the SPICA/SAFARI wavelength
range.

Real galaxies will of course produce a mixture
of PDR and XDR emission. Separating these will require (a) a complete inventory
of the cooling emission lines at high sensitivity; (b) accurate modeling with
state-of-the-art PDR and 
XDR codes, including radiative transfer in 3 dimensions;
(c) additional data for constraining local densities (density-tracing
molecules), column densities (isotopomeric lines), sizes of emission regions
(high resolution ALMA imaging), as well as high-quality X-ray and infrared data.
All of these will be in reach within the coming decade, and thus we may
have some optimism that a reliable
separation of star formation and AGN as contributors
to the total luminosity of galaxies (which has been elusive problem for
many years) can indeed be achieved with SPICA\null.

\begin{figure}[th]
  \begin{center}
    \includegraphics[width=8cm]{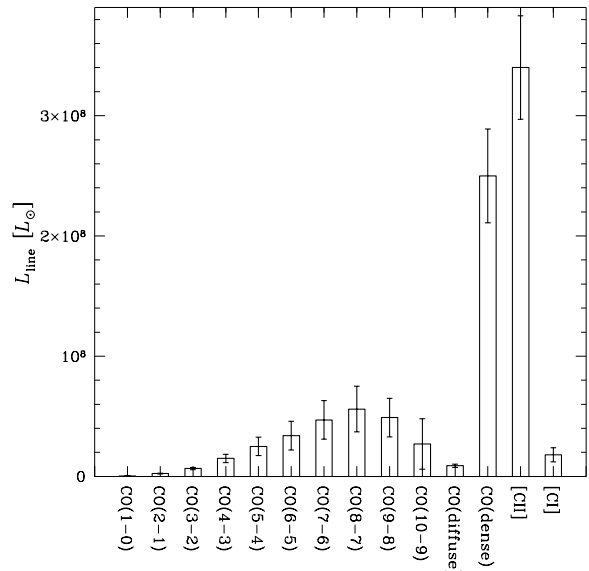}
  \end{center}
  \caption{Cooling budget in \object{Mrk\,231} \citep{Papadopoulosetal07},
based on observations until CO $J=6{-}5$ and a well-constrained model for the
higher lines. The CO is modeled as a combination of a diffuse and a dense phase,
with the dense phase containing most of the mass and emitting most of the
cooling radiation, comparable in total luminosity to the 
[C\,II] line.}
\label{vanderWerfP_fig2}
\end{figure}

\section{State-of-the-art: low-$z$ galaxies}
\label{vanderWerfP.seclowz}

Reasonably detailed CO spectral line energy distributions are
gradually becoming available, for instance for M82 \citep{Maoetal00} and
NGC\,253 \citep{Bradfordetal03,Gustenetal06},
although for low-$z$ galaxies they still
do not probe higher rotational levels than $J=7$. The first
high-quality CO flux distribution for a ULIRG was presented by 
\citet{Papadopoulosetal07} for the nearby ULIRG \object{Mrk\,231}. These authors
conclude that the CO ladder requires
a two-component model: one component of fairly
diffuse gas dominating the low-$J$ lines and one component of dense gas
dominating at $J>4$. Remarkably, it was found that the dense component dominates
the molecular gas mass as well as the total cooling radiation from
\object{Mrk\,231}.
The total cooling by this component approaches that by the [\cii] line,
implying a totally different thermal balance than in for instance the
Milky Way. The derived cooling budget is shown in Fig.~\ref{vanderWerfP_fig2}.
This result also sheds new light on the
discovery with the ISO satelite that in ULIRGs the
[\cii]/$\LIR$ ratio is about a factor of 10 lower than in lower
luminosity galaxies (where the [\cii] $158\,\mu$m line typically
contains $0.1-0.5$\% of $\LIR$).
Physically, this situation can arise in dense PDRs (where
the ionized carbon layer is thin and the molecular gas layer is close
to the heating source), in XDRs, or in PDRs with an enhanced cosmic ray flux.
These possibilities can be distinguished using the more comprehensive data that
will be obtained by the Herschel Comprehensive ULIRG Emission Survey
(HerCULES). This
project will obtain full spectra of the CO ladder from $J=5$ to 13 as well as
the key cooling lines of [\cii] ($158\,\mu$m) and [\oi] (63 and $145\,\mu$m)
for a flux-limited sample of 29 local LIRGs and ULIRGs. This project will be
an important pathfinder for future studies with SPICA/SAFARI\null. 

\begin{figure}[th]
  \begin{center}
    \includegraphics[width=8cm]{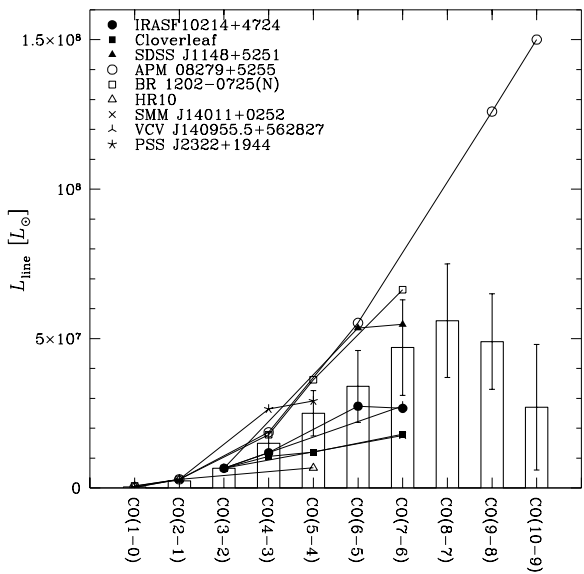}
  \end{center}
  \caption{CO  line   luminosities  in  various   high-redshift  objects
  compared to the Mrk\,231 template (bar diagram).
Line luminosities are normalized
  to   the  Mrk\,231   values   at  either   CO   $J{=}3{-}2$  or   CO
  $J{=}2{-}1$ \citep{Papadopoulosetal07}.}
\label{vanderWerfP_fig3}
\end{figure}

\section{State-of-the art: high-$z$ galaxies}

To date, detections of CO have been reported in about 50 objects at
$z>1$, 38 of which are mentioned in the review by \citet{SolomonVandenBout05}.
Of these 38, 22 are at redshifts $z=1.5-2.8$, the time in
cosmic evolution in which both
the QSO and the SMG activity peak. Multi-line CO studies of the
12 brightest of these (8~QSOs and 4~SMGs with no signs of AGN activity)
are being carried out using the IRAM and APEX telescopes. These
observations typically cover CO rotational lines from $J=3{-}2$ to
$J=8{-}7$ \citep{Weissetal05}. A compilation, compared with the
\object{Mrk\,231} flux distribution of \citet{Papadopoulosetal07} is shown in
Fig.~\ref{vanderWerfP_fig3}. 

The location of the peak of the CO flux distribution
reveals different excitation conditions within the
sample.  For QSOs the peak typically occurs at the
$J=6{-}5$ or $7{-}6$ transition, indicative of a gas density of
$\sim10^4\,$cm$^{-3}$ and a kinetic temperature of $\sim50\,$K\null.
For \object{APM\,0827+5255} the CO excitation is by higher by a large margin.
Here the flux distribution peaks at the $J=10{-}9$
transition. The SMGs in the sample tend to have a lower gas excitation
than the QSOs with a peak of the CO ladder typically at the $J=5{-}4$
line.  These CO ladders are similar to those observed in the central
region of some local starburst galaxies (e.g., M82 and
NGC\,253, see references in Section~\ref{vanderWerfP.seclowz}). 
The lowest CO excitation is
observed toward the
extremely red object \object{HR\,10}
where the turnover of the flux distribution already
occurs at the CO $J=4{-}3$ line \citep{PapadopoulosIvison02}.

Although this set of CO lines constrains
the properties of the bulk of the CO emitting gas, surprising 
results are obtained when the CO ladder is probed beyond the peak,
as can be seen in the
two QSOs at $z\sim4$
(\object{APM\,08279+5255} and \object{PSS\,2322+1944}) where
the observations even include CO lines up to 
$J=11{-}10$, in the 
rest-frame THz regime \citep{Weissetal07}.
The CO $J=9{-}8$ and $11{-}10$ lines in the QSO host galaxy
\object{PSS\,2322+1944}
($z=4.1$) show a clear flux excess compared to the single
excitation model which describes the peak of the CO ladder
and require a second gas component with even higher CO excitation.
This component may be related to a denser component (also
indicated by HCN observations), or alternatively to a hotter
component heated by the AGN such as in \object{APM\,08278+5255}, and can be
modeled using the CO ladders shown in Fig.~\ref{vanderWerfP_fig1}.
Interestingly, a similar
CO ladder is observed towards the \object{Galactic center}, but there
the high-excitation component is energetically much less important.

In summary, observations of high-$z$ sources allow access to diagnostic
lines of CO which are very difficult to obtain from ground
based observations for sources in the local universe. This leads to
the somewhat paradoxical situation, that the gas properties for some high-$z$
redshift objects are currently better constrained than those of most
nearby infrared-luminous galaxies. The situation will be significantly improved
by the HerCULES project, referred to in Section~\ref{vanderWerfP.seclowz}. Expanding
such observations
to lower luminosity galaxies, larger distances and higher $J$ CO lines, in order
to fully exploit the diagnostic use of the CO ladder,
will be a major task for SPICA/SAFARI\null.

\begin{figure}[th]
  \begin{center}
    \includegraphics[width=8cm]{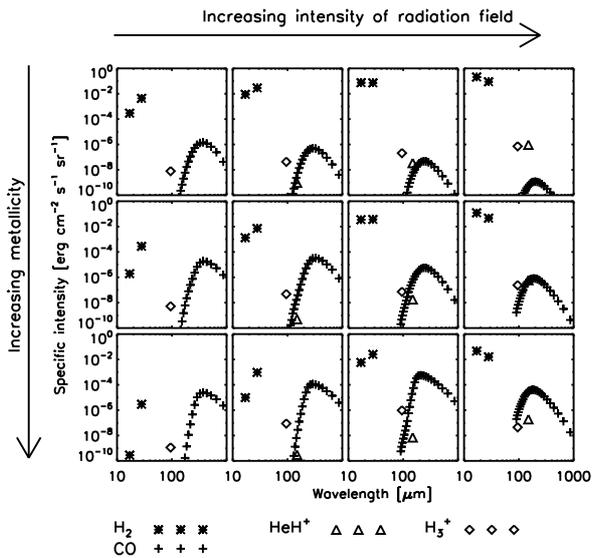}
  \end{center}
  \caption{Rest frame spectral line distributions of high-$J$ CO, H$_2$
S(0) and S(1) (as well as lines from H$_3^+$
and HeH$^+$) are shown as functions of metallicity and X-ray flux for
a density of $10^5\,$cm$^{-3}$. The impinging X-ray flux takes on values of
0.1, 1, 10 and 100\,erg\,s$^{-1}$\,cm$^{-2}$; and the
metallicity values of $10^{-3}$, $10^{-2}$ and $10^{-1}$ of solar
\citep{SpaansMeijerink08}.}
\label{vanderWerfP_fig4}
\end{figure}

\section{H$_2$ cooling emission from primordial galaxies}

A key question is what happens in galaxies of extremely low metallicity,
where the [\cii], [\oi] and CO cooling channels are not available.
In standard cosmological models, a significant population of dark matter
halos collapses at redshifts larger than 5, aided by atomic (H\,{\sc i} and
He\,{\sc ii}
Lyman $\alpha$) and molecular (H$_2$ and HD) cooling, and hosts the first
stars and accreting black holes \citep{Haimanetal00}. The H$_2$
rotational and ro-vibrational lines are accessible to SAFARI and cool the
(molecular) medium in systems with metallicities $\le 10^{-2}-10^{-3}$
Solar. As shown by \citet{Mizusawaetal05} and \citet{SpaansMeijerink08},
a number of these lines are expected to be detectable with JWST, ALMA and
SPICA/SAFARI, over a wide range of physical conditions.

H$_2$ cooling emission from population~III star forming regions at
$z\sim 20$ will be too faint for detection, but primordial
star forming gas in $z\sim 5$ galaxies should be detectable with
SAFARI due to the enhanced ionization and molecule formation caused by
virialization shocks \citep{Mizusawaetal05}.  Furthermore, the
presence of an accreting black hole leads to the copious production of
X-rays. In such X-ray irradiated gas, typical line intensities can be
of the order of $0.1-1.0\,$erg s$^{-1}$ cm$^{-2}$ sr$^{-1}$ for the pure
rotational lines of H$_2$ at 28, 17, etc.\ $\mu$m, which is as much as
$10-30\%$ of the black hole's accretion-driven X-ray luminosity
\citep{SpaansMeijerink08}. Flux distributions
for various parameters are shown in
Fig.~\ref{vanderWerfP_fig4}, which shows the dominant role of H$_2$ at the lowest
metallicities.

Given the tight relation between supermassive black holes and the
stellar spheroids that host them, 
it seems reasonable to use the extended Press-Schechter
formalism and assume that (a) the $M_\bullet-\sigma$ holds at high
redshift too, so the black hole mass is $2\times 10^{-4}$ of the total
dark matter mass, that (b) black holes radiate at 10\% of Eddington,
yielding an X-ray dominated zone of $50-400$ pc, and that (c) star
formation and black hole growth are roughly simultaneous, with similar
contributions to the luminosity. The latter implies star formation rates
of $10^2\,M_\odot$\,yr$^{-1}$ within the central 2\,kpc, quickly enriching
the pristine gas, but the presence of a black hole alone is quite
sufficient for SAFARI detections.

For the $\Lambda$CDM concordance model, one then finds about 10 of such
sources per $2'\times 2'$ field-of-view. 
The main contribution comes from $z<10$
halos with masses $\ge 10^9$ M$_\odot$ that yield H$_2$ line fluxes of
$\ge 10^{-18}$ W m$^{-2}$, comfortably above the 
$2\times 10^{-19}$ W\,m$^{-2}$ SAFARI detection threshold.
Typically, at least two pure
rotational H$_2$ lines (e.g., S(0), S(1), S(2), etc.) are detectable for
a $z\sim 5$ source, allowing line confusion due to redshifted mid-infrared
lines from
far-infrared sources to be dealt with. SPICA/SAFARI targets will likely
be provided by JWST, which will discover many $z\ge 5$ galaxies.

\begin{acknowledgements}
Much of the work reported here was carried out in collaboration with
Padelis Papadopoulos, Kate Isaak and Rowin Meijerink, without
whose enthusiasm and perseverance these projects would have been
less successful and less fun. We also thank Dominik Schleicher and Ralf Klessen
for good discussions.
\end{acknowledgements}

\end{document}